\documentclass[10pt,letterpaper,twocolumn,aps,pra,showpacs,superscriptaddress,longbibliography]{revtex4-1}
\pdfoutput=1

\usepackage{layouts}
\usepackage{physics}
\usepackage{ucs}
\usepackage{amsmath}
\usepackage{amsfonts}
\usepackage{amssymb}
\usepackage{makeidx}
\usepackage{cellspace,booktabs}
\usepackage{natbib}
\usepackage{hyperref}
\usepackage[capitalise]{cleveref} 
\usepackage{xcolor}
\usepackage{microtype}
\usepackage{bm}
\usepackage{graphicx}
\usepackage{multirow}

\renewcommand{\dag}{^{\dagger}}
\newcommand{\exv}[1]{ \left\langle #1 \right\rangle }

\begin{document}
\date{\today}

\author{Christian~Kraglund~Andersen}
\thanks{These authors contributed equally to this work.}
\affiliation{Department of Physics, ETH Zurich, CH-8093 Zurich, Switzerland}
\author{Ants~Remm}
\thanks{These authors contributed equally to this work.}
\affiliation{Department of Physics, ETH Zurich, CH-8093 Zurich, Switzerland}
\author{Stefania~Balasiu}
\affiliation{Department of Physics, ETH Zurich, CH-8093 Zurich, Switzerland}
\author{Sebastian~Krinner}
\affiliation{Department of Physics, ETH Zurich, CH-8093 Zurich, Switzerland}
\author{Johannes~Heinsoo}
\affiliation{Department of Physics, ETH Zurich, CH-8093 Zurich, Switzerland}
\author{Jean-Claude~Besse}
\affiliation{Department of Physics, ETH Zurich, CH-8093 Zurich, Switzerland}
\author{Mihai~Gabureac}
\affiliation{Department of Physics, ETH Zurich, CH-8093 Zurich, Switzerland}
\author{Andreas~Wallraff}
\affiliation{Department of Physics, ETH Zurich, CH-8093 Zurich, Switzerland}
\author{Christopher~Eichler}
\affiliation{Department of Physics, ETH Zurich, CH-8093 Zurich, Switzerland}

\title{Entanglement Stabilization using Parity Detection\\and Real-Time Feedback in Superconducting Circuits}

\date{\today}

\begin{abstract}
Fault tolerant quantum computing relies on the ability to detect and correct errors, which in quantum error correction codes is typically achieved by projectively measuring multi-qubit parity operators and by conditioning operations on the observed error syndromes. Here, we experimentally demonstrate the use of an ancillary qubit to repeatedly measure the $ZZ$ and $XX$ parity operators of two data qubits and to thereby project their joint state into the respective parity subspaces. By applying feedback operations conditioned on the outcomes of individual parity measurements, we demonstrate the real-time stabilization of a Bell state with a fidelity of $F\approx74\%$ in up to 12 cycles of the feedback loop. We also perform the protocol using Pauli frame updating and, in contrast to the case of real-time stabilization, observe a steady decrease in fidelity from cycle to cycle. The ability to stabilize parity over multiple feedback rounds with no reduction in fidelity provides strong evidence for the feasibility of executing stabilizer codes on timescales much longer than the intrinsic coherence times of the constituent qubits.
\end{abstract}
\maketitle

The inevitable interaction of quantum mechanical systems with their environment renders quantum information vulnerable to decoherence~\cite{Gardiner2004, Haroche2006, Schlosshauer2007}. Quantum error correction aims to overcome this challenge by redundantly encoding logical quantum states into a larger-dimensional Hilbert space and performing repeated measurements to detect and correct for errors~\cite{Shor1995, Steane1996a, Lidar2013, Terhal2013}. For sufficiently small error probabilities of individual operations, logical errors are expected to become increasingly unlikely when scaling up the number of physical qubits per logical qubit~\cite{Raussendorf2007, Fowler2012}. As the concept of quantum error correction provides a clear path toward fault tolerant quantum computing \cite{Gottesman2010}, it has been explored in a variety of physical systems ranging from nuclear magnetic resonance~\cite{Cory1998}, to trapped ions~\cite{Chiaverini2004, Schindler2011, Lin2013b, Barreiro2011} and superconducting circuits, both for conventional \cite{Reed2012, Shankar2013a, Riste2015, Kelly2015} and for continuous variable based encoding schemes~\cite{Ofek2016}.

Quantum error correction typically relies on the measurement of a set of commuting multi-qubit parity operators, which ideally project the state of the data qubits onto a subspace of their Hilbert space -- known as the code space -- without extracting information about the logical qubit state \cite{Terhal2013}. A change in the outcome of repeated parity measurements signals the occurrence of an error, which brings the state of the qubits out of the code space. Such errors can either be corrected for in real-time by applying conditional feedback, or by keeping track of the measurement outcomes in a classical register to reconstruct the quantum state evolution in post-processing. The latter approach -- also known as Pauli frame updating~\cite{Knill2005} -- has the advantage of avoiding errors introduced by imperfect feedback and additional decoherence due to feedback latency. Real-time feedback, on the other hand, could be beneficial in the presence of asymmetric relaxation errors, by preferentially mapping the qubits onto low energy states, which are more robust against decay~\cite{OBrien2017}. There are also important instances in which knowledge of the measurement results is required in real-time to correctly choose subsequent operations, e.g., for the realization of logical non-Clifford gates~\cite{Terhal2013}, for measurement-based quantum computing~\cite{Raussendorf2001, Briegel2009}, and for stabilizing quantum states in cavity systems~\cite{Sayrin2011, Ofek2016}. Therefore, parity measurements, conditional feedback and Pauli frame updating are all important elements for fault tolerant quantum computing and are explored very actively.

Parity detection has previously been studied with superconducting circuits both with and without the use of ancillary qubits. Joint dispersive readout~\cite{Lalumiere2010, Riste2013} and the quantum interference of microwave signals~\cite{Roch2014,Roy2015} were used to deterministically generate Bell states and their stabilization was achieved by autonomous feedback based on reservoir engineering~\cite{Shankar2013a}. Moreover, a recent theortical proposal presented a protocol for direct weight-4 parity detection~\cite{Royer2018}. However, the most common approach for parity detection uses an ancillary qubit onto which the parity of the data qubits is mapped and then projectively measured by reading out the ancillary state. By using two ancillary qubits, the simultaneous measurement of the $XX$ and $ZZ$ parity operators of two data qubits was demonstrated~\cite{Corcoles2015}. Furthermore, ancilla based parity detection has enabled the realization of a three-qubit bit flip code~\cite{Riste2015}, a five qubit repetition code~\cite{Kelly2015}, and the measurement of multi-qubit parity operators for three~\cite{Blumoff2016} and four~\cite{Takita2016} data qubits. Repeated parity detection was also achieved for the cat code~\cite{Sun2014}.

In most previous implementations of ancilla-based parity detection, changes in the measured parity were accounted for in post-processing rather than actively compensated for using feedback. Conditional feedback, however, was previously used in superconducting circuits to initialize and reset qubit states~\cite{Riste2012,Salathe2018}, to demonstrate a deterministic quantum teleportation protocol~\cite{Steffen2013}, and to extend the lifetime of a qubit state encoded as a cat state in a superconducting cavity~\cite{Ofek2016}.

\begin{figure}[b]
\includegraphics[width=\linewidth]{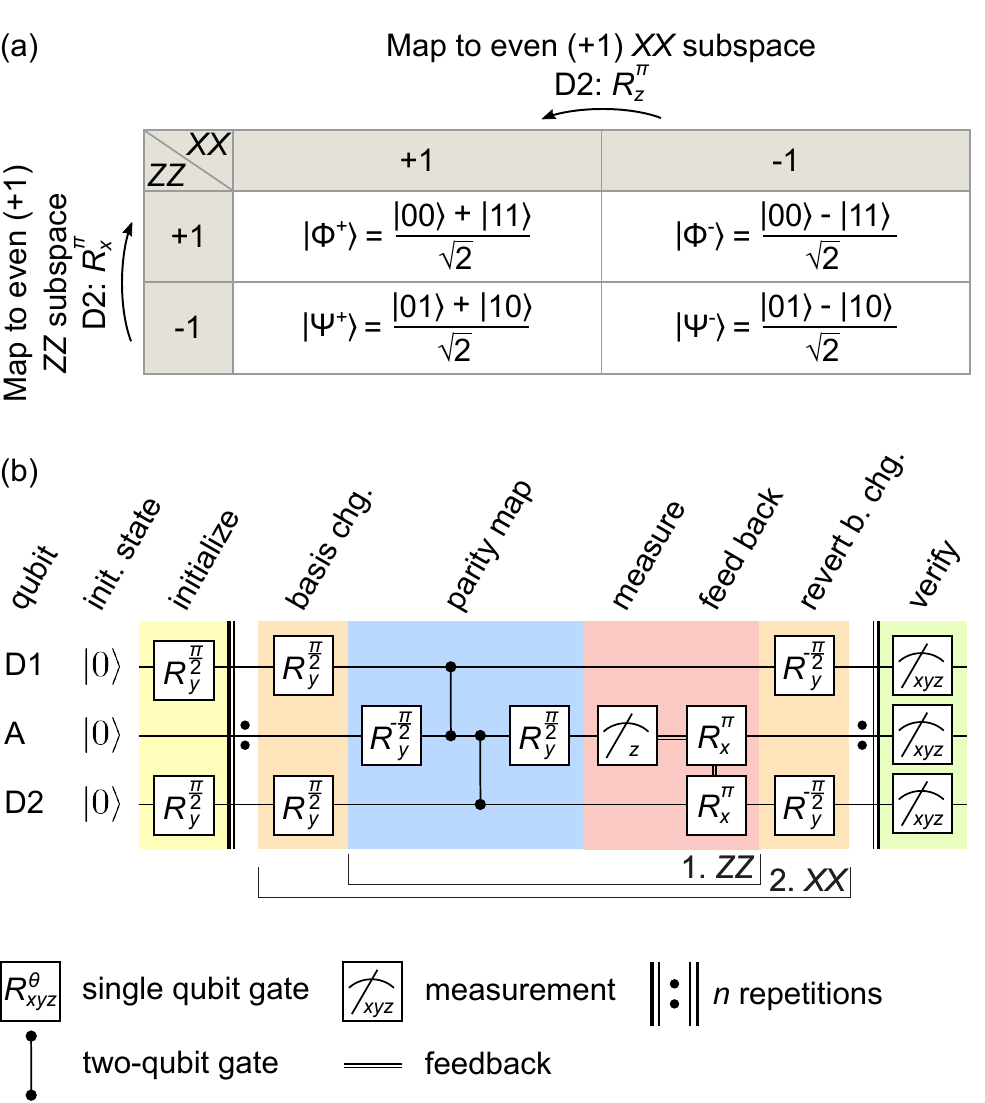}
\caption{(a) Joint eigenstates of the parity operators $ZZ$ and $XX$ for the four different combinations of eigenvalues. Mapping to the target state $|\Phi^+\rangle$ is achieved by applying a $\pi$-rotation to qubit $D2$ around the $x$-axis ($z$-axis), if $ZZ$ ($XX$) is measured to be $-1$. (b) Gate sequence of the parity stabilization protocol which deterministically projects the qubits onto a unique Bell state. Vertical lines with dots represent conditional phase gates, $R_y^{\theta}$ are rotations about the $y$ axis by an angle $\theta$. Conditional feedback operations are indicated by a double line connected to a measurement operation. The vertical bars indicate a repetition of either the $ZZ$, or the combined $ZZ$ (1.) and $XX$ (2.) parity detection and feedback.
}
\label{fig:1}
\end{figure}

Here, we report on the experimental realization of repeated $XX$ and $ZZ$ parity detection of two superconducting qubits. In contrast to previous experiments in superconducting circuits, we perform real-time conditional feedback to stabilize the data qubits in a Bell state and to actively reset the ancillary qubit to the ground state, see Fig.~\ref{fig:1}. Our results are, thus, closely related to the recent experiments realized in a trapped ion system~\cite{Negnevitsky2018}. We note that similar experiments using Pauli frame updating rather than real-time feedback have been performed in parallel with our work~\cite{Bultink2019}.

\begin{figure*}[t]
\includegraphics[width=\linewidth]{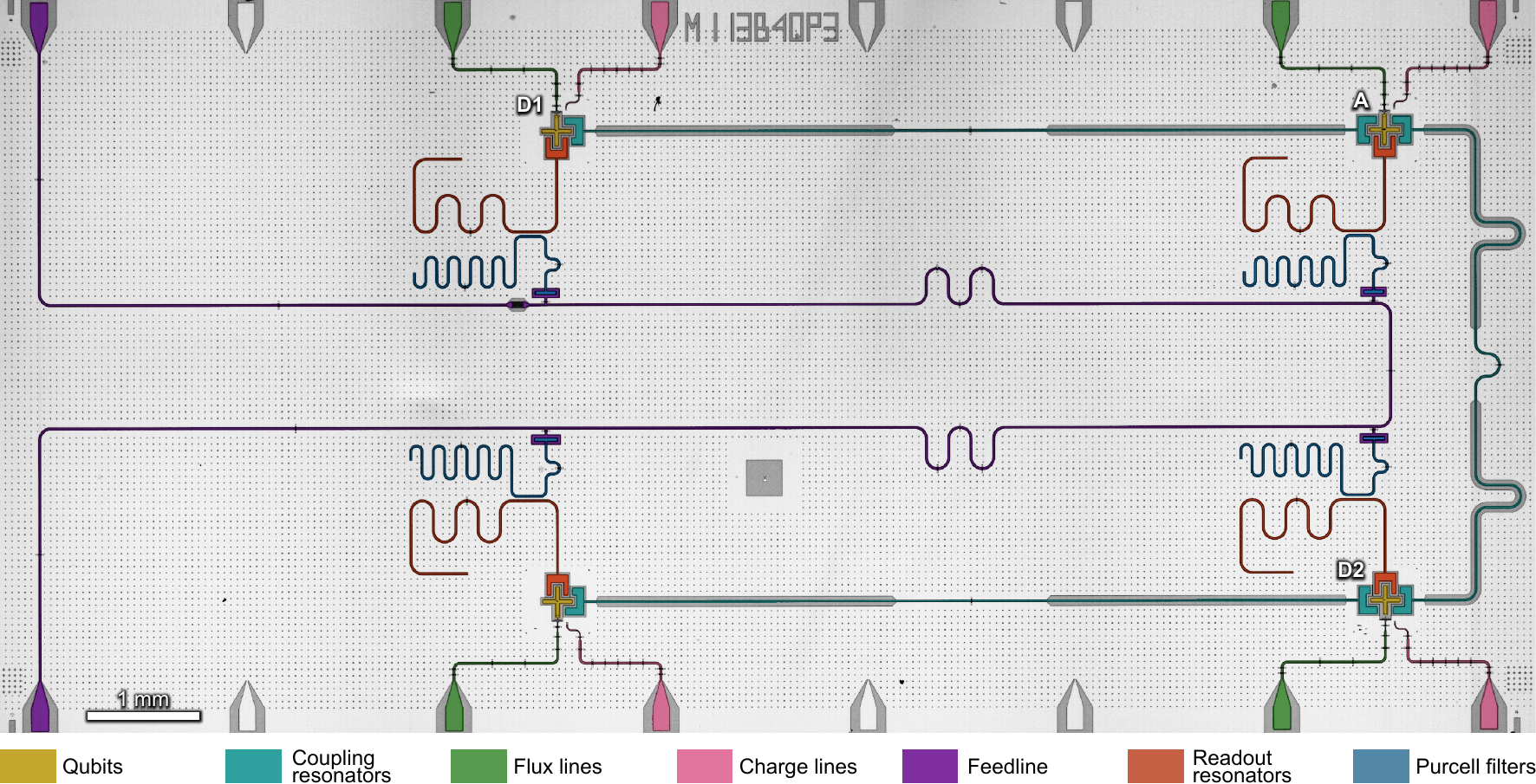}
\caption{
False colored micrograph of the four-qubit device used in this work, with transmon qubits shown in yellow, coupling resonators in cyan, flux lines for single-qubit tuning in  green, charge lines for single-qubit manipulation in pink, a common feedline for readout in purple, and transmission line resonators for readout and for Purcell filtering in red and blue, respectively.
}
\label{fig:2}
\end{figure*}

The objective of the protocol is to stabilize two data qubits $D1$ and $D2$ in a target Bell state, chosen to be $|\Phi^+\rangle=(|00\rangle+|11\rangle)/\sqrt{2}$, for which both the $XX$ and the $ZZ$ parity are even, i.e. take the value $+1$, see Fig.~\ref{fig:1}(a). We initially prepare both data qubits in an equal superposition state by applying a rotation $R_y^{\theta}$ to both qubits with an angle $\theta=\pi/2$ around the ${y}$-axis. We then map the parity of the joint state of $D1$ and $D2$ onto the ancillary qubit $A$ by applying two controlled NOT gates -- decomposed into conditional phase gates~\cite{DiCarlo2009} and single qubit rotations -- with the ancillary qubit as the target controlled by each of the data qubits. The subsequent measurement of $A$ probabilistically yields the measurement result $\ket{0}$ ($\ket{1}$) for the ancilla qubit state, indicating the parity operator eigenvalue $+1$ ($-1$) and, ideally, projecting the joint state of $D1$ and $D2$ into the corresponding even (odd) parity subspace spanned by the basis states $|{00}\rangle$ and $|11\rangle$ ($|{01}\rangle$ and $|{10}\rangle$). The probability for each outcome depends on the input state and is ideally $50\%$ when initializing both qubits in an equal superposition state. After the measurement of $A$, we map the state of the data qubits into the even parity subspace by flipping the state of $D2$ with a $R^{\pi}_x$ pulse if the parity measurement yields $-1$~\cite{Riste2012}. In this case, we also reset the ancillary qubit to the ground state in preparation for the next round of parity detection.

Similarly, we perform $XX$ parity measurements by changing the basis of $D1$ and $D2$ before and after the parity stabilization sequence with $R^{\pm\pi/2}_x$-pulses, see Fig.~\ref{fig:1}(b). By choosing a feedback protocol that stabilizes both the $XX$ and the $ZZ$ parity to be even, we project the two data qubits onto the unique Bell state $|\Phi^+\rangle=(|00\rangle + |11\rangle)/{\sqrt{2}}$ in two subsequent rounds of parity feedback. Repeating these two parity stabilization steps sequentially ideally stabilizes this Bell state indefinitely. The main requirements for the realization of this protocol are (i) high fidelity and fast readout of the ancillary qubit with little disturbance of the data qubits, (ii) high fidelity single- and two-qubit gates, (iii) low latency classical electronics to perform conditional feedback with delay times much shorter than the qubit coherence times, and (iv) the absence of leakage into non-computational states.

We implement this parity stabilization protocol on a small superconducting quantum processor consisting of a linear array of four transmon qubits, of which each pair of nearest neighbors is coupled via a detuned resonator~\cite{Majer2007}, see Fig.~\ref{fig:2}. Each qubit has individual charge (pink) and flux control  lines (green) to perform single qubit gates and to tune the qubit transistion frequency.  Each of the four qubits is coupled to an individual readout circuit used for probing the state of the qubits by frequency multiplexed dispersive readout through a common feedline (purple)~\cite{Heinsoo2018}. Further details of the device and its fabrication are discussed in Appendix~\ref{app:sample}. We mount this four-qubit device at the base plate of a cryogenic measurement setup, equipped with input and output lines, specified in Appendix~\ref{app:setup}, for microwave control and detection.

For our experiments we use the three qubits labeled $D1$, $A$, and $D2$ in Fig.~\ref{fig:2}. We perform single qubit gates in $50\,$ns using DRAG pulses~\cite{Motzoi2009} with an average gate error of 0.3\% characterized by randomized benchmarking~\cite{Magesan2011}. Single qubit gate fidelities are mostly limited by the coherence times, which range from $14$ to $23\,\mu s$ (15 to 22$\,\mu$s) for $T_1$ ($T_2^E$), see Table~\ref{tab:qb_params} in Appendix~\ref{app:sample}. Conditional phase gates are realized with flux pulses on the data qubits in approximately 180$\,$ns, by tuning the $|11\rangle$ state into resonance with the $|20\rangle$ state for a full period of the resonant exchange interaction~\cite{DiCarlo2009}, see also Appendix~\ref{app:setup}. By calibrating and correcting for flux pulse distortion, we achieve two-qubit gate fidelities of approximately 99\% characterized using quantum process tomography~\cite{Nielsen2000}. The dynamical phases acquired during the flux pulses are compensated by using virtual-$Z$ gates~\cite{McKay2017}. Using a readout pulse length of 200~ns and an integration time of 400~ns, see Appendix~\ref{app:readout}, we achieve an average probability for correct readout assignment close to 99\%, when reading out all three qubits simultaneously.

\begin{figure*}[t]
\includegraphics[scale=1]{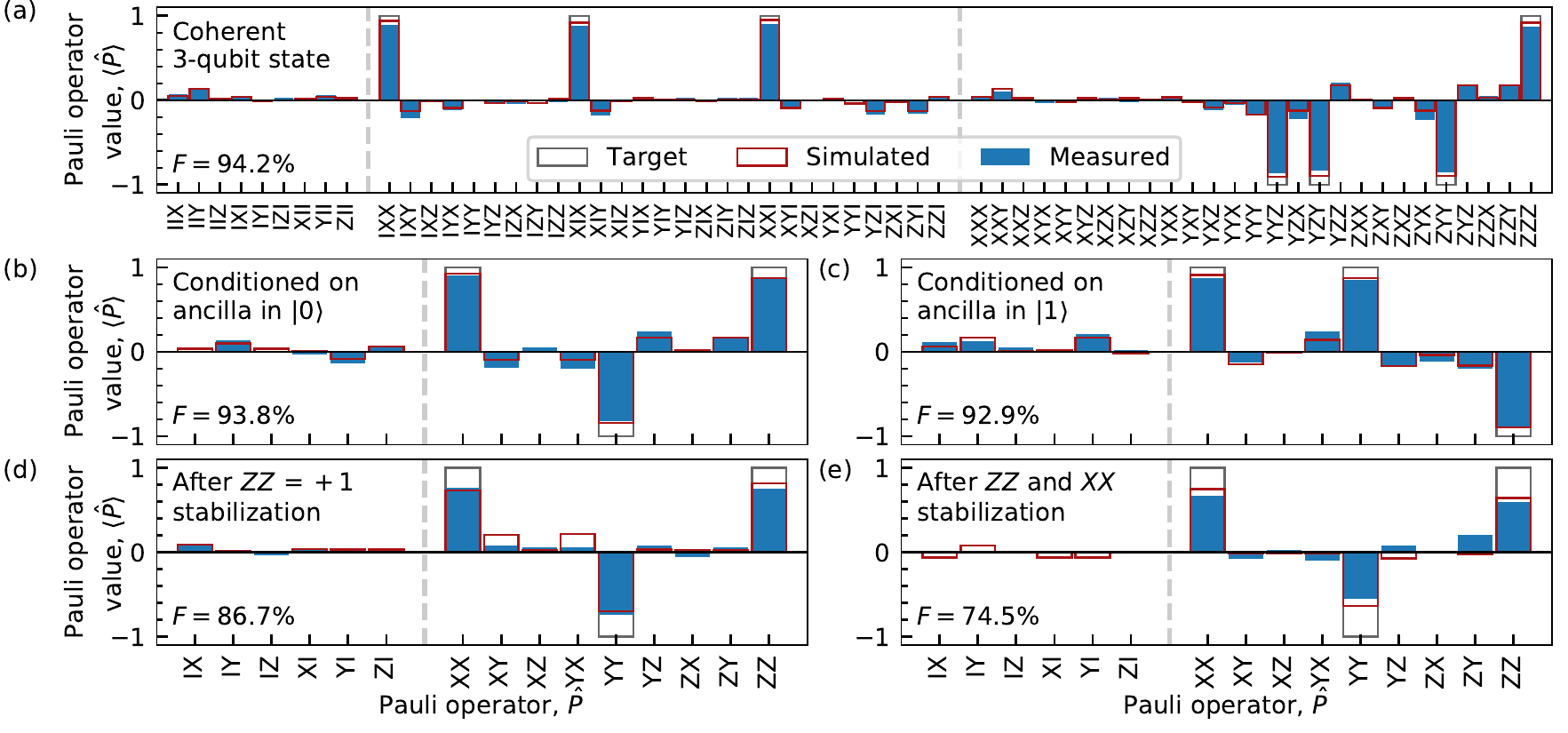}
\caption{Ideal (black frame), simulated (red frame), and measured (blue) expectation values of multi-qubit Pauli operators $\hat{P}$ for (a) the 3-qubit state prior to the first ancilla readout in the basis of $D1$, $A$, $D2$, (b)-(c) the state of $D1$ and $D2$ conditioned on the respective measurement outcome of $A$, when simultaneously measuring all three qubits, (d) the unconditional state after the first $ZZ$ parity measurement and conditional feedback, and (e) after one round of consecutive $ZZ$ and $XX$ parity stabilization. The quoted fidelities are calculated from the most likely density matrix reconstructed based on the measured Pauli sets. Vertical dashed lines separate the single-qubit operators from two-qubit (and three-qubit) correlations.
}
\label{fig:3}
\end{figure*}

\begin{figure}[t]
\includegraphics[width=\linewidth]{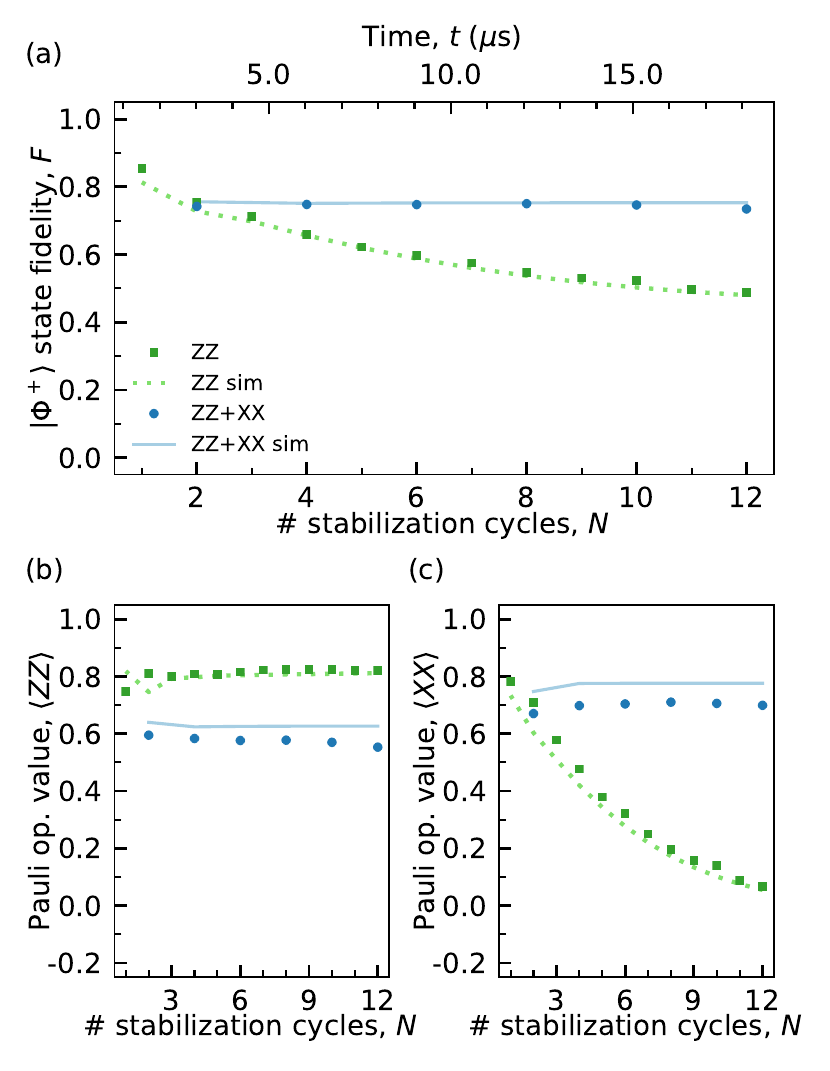}
\caption{
(a) Measured (dots and squares) and simulated (solid and dashed lines) $|\Phi^+\rangle$ state fidelity $F$ after $N$ rounds of $ZZ$ parity stabilization (green), and when iterating between $ZZ$ and $XX$ (blue). (b) - (c) Measured and simulated expectation values $\langle ZZ \rangle$ and $\langle XX\rangle$ for the same two experiments.
}
\label{fig:4}
\end{figure}

We characterize the coherent part of the parity detection algorithm by performing quantum state tomography of all three qubits prior to the first readout of $A$. Accounting for finite readout fidelity, we average this data to obtain the expectation values for all multi-qubit Pauli operators $\hat{P}$ [Fig.~\ref{fig:3}(a)]. The overall three-qubit state fidelity $F = \langle \psi|\rho|\psi\rangle \approx 94\%$, estimated based on the most likely density matrix $\rho$ reconstructed from the measured Pauli sets, is in good agreement with the fidelity of $92.8\%$, calculated using a master equation simulation accounting for qubit decoherence and residual $ZZ$ coupling (for details see Appendix~\ref{app:simulations}). The finite $XY$ correlations, which are well reproduced by the numerical simulations, are due to the residual $ZZ$ coupling between the data qubits and the ancillary qubit with rates 110 kHz and 370 kHz for D1 and D2, respectively, which we do not compensate for during the coherent part of the protocol. A reduction of residual $ZZ$ coupling could, e.g., be achieved with alternative coupling schemes featuring larger on-off ratios~\cite{McKay2015, Yan2018b, Zhang2018k}.

In a next step, we characterize the state of $D1$ and $D2$ conditioned on the outcome of the first ancilla measurement using two-qubit state tomography, see Fig.~\ref{fig:3}(b)-(c). In this experiment, both $D1$ and $D2$ are read out simultaneously with $A$. For both parity measurement outcomes +1 and $-1$, projecting $D1$ and $D2$ into the even or odd parity Bell state $|\Phi^+\rangle$ and $|\Psi^+\rangle$, respectively, we find the resulting fidelities (93.8\% and 92.9\%) to be close to the ones obtained by projecting the reconstructed three-qubit state onto the corresponding two-qubit subspaces (95.9\%  and 93.4\%). This level of agreement is consistent with the high readout assignment probability of $98.7\%$ of the ancillary qubit. Most importantly, we find the outcome of the ancilla measurement to correlate very well with the sign of the resulting $ZZ$ correlations of the data qubits, indicating that the parity measurement is highly projective. More specifically, we find $\langle ZZ \rangle = +0.86$ $(-0.89)$ conditioned on having measured $A$ in the ground (excited) state.

To prepare the specific target state $|\Phi^+\rangle = (|00\rangle + |11\rangle)/{\sqrt{2}}$ deterministically, we apply a $\pi$-pulse to qubit $D2$ if the ancilla measurement yields an odd parity $-1$, compare circuit diagram in Fig.~\ref{fig:1}(b). Alternatively, we could prepare the state $|\Psi^+\rangle =(|10\rangle + |01\rangle)/{\sqrt{2}}$, by changing the condition for feedback, i.e., applying the feedback pulse if the parity is even. The feedback scheme chosen here, maps the odd parity state characterized in Fig.~\ref{fig:3}(c) to the even parity state in (b). Indeed, the resulting unconditional state has a fidelity of 86.7\% indicating that we correctly prepare the desired target state, see Fig.~\ref{fig:3}(d). According to the comparison with master equation simulations (see Appendix \ref{app:simulations} for details), the reduced fidelity is dominated by qubit decoherence during the delay time of $1\,\mu$s between the parity detection and the application of the feedback pulse to $D2$ (Appendix~\ref{app:setup}). We partly mitigate the dephasing and the residual $ZZ$ interaction by applying four dynamical decoupling pulses using the Carr-Purcell-Meiboom-Gill (CPMG) protocol to the data qubits during the feedback delay time, see pulse sequence in Appendix~\ref{app:setup}. We observe deterministic phase shifts of both data qubits after completion of the ancilla readout, which we attribute to a measurement induced Stark shift due to the off-resonant driving of coupling resonators by the ancilla readout pulse~\cite{Pechal2012}. We compensate for these phase shifts by inserting virtual Z gates after the ancilla readout. However, we possibly over-correct this source of error and, thus, observe smaller phase errors ($\exv{XY}$ and $\exv{YX}$) in the experiment than expected from simulations [Fig.~\ref{fig:3}(d)].

We emphasize that the $XX$ and $YY$ correlations, measured after the $ZZ$-parity check [Fig.~\ref{fig:3}(d)], are a consequence of the specific initial state $(|00\rangle+|01\rangle+|10\rangle+|11\rangle)/2$, which we prepare prior to mapping onto the even $ZZ$ parity subspace. For the more general case of, e.g.,~an mixed initial state, we observe a nearly vanishing $\exv{XX}$ correlation of $0.011$ after mapping onto the even $ZZ$ subspace, as expected. Creating and stabilizing finite $XX$ [and $YY = -(XX)(ZZ)$] correlations, therefore requires a  consecutive measurement of the commuting $XX$ parity operator and the projection of $D1$ and $D2$ into the corresponding subspace. We achieve this by enclosing the $ZZ$ parity stabilization pulse sequence in appropriately chosen basis change rotations, see Fig.~\ref{fig:1}(b). The resulting state has a fidelity of $74.5\%$ compared to the target state $|\Phi^+\rangle$ [Fig.~\ref{fig:3}(e)], close to the simulated value of $75.8\%$. From simulations, we find that the reduction in fidelity relative to the previous round of parity feedback is dominated by the additional dephasing of data qubits during the $XX$ stabilization cycle.

Most importantly, we also demonstrate the repeatability of parity detection and stabilization which is a crucial requirement for quantum error correction. Specifically, we characterize the evolution of the prepared quantum state for up to 12 cycles of $ZZ$ or $XX$ parity stabilization sequences. We first repeatedly measure the $ZZ$ parity and stabilize the state in the even $ZZ$ subspace. In this case, we observe a decrease of the measured Bell state fidelity  to $\sim50\%$ after $N=12$ cycles [green points in Fig.~\ref{fig:4}(a)]. This experimental observation is in agreement with master equation simulations and due to the decoherence of the initial $\langle XX \rangle$ correlations, which are not stabilized by the repeated $ZZ$ parity checks, see green symbols in Fig.~\ref{fig:4}(c). When interleaving the $ZZ$ parity stabilization with $XX$ parity stabilization, we observe the expected unconditional stabilization of the target Bell state. Already after a single pair of stabilization cycles ($N=2$), the Bell state fidelity reaches a steady state value of $\sim 74\%$, which is maintained for all subsequent stabilization cycles.

To gain further insights into the feedback process, we also perform the experiment using Pauli frame updates rather than applying feedback pulses to the data qubits while still using feedback for resetting the ancilla qubit, see Appendix~\ref{app:pauli} for details. In this case, we observe a $10\%$ lower Bell state fidelity after $N=12$ cycles, half of which is expected from simulations. We attribute the decrease in fidelity to the asymmetry of relaxation errors~\cite{OBrien2017}. In the absence of feedback, the data qubits remain in the odd $ZZ$ subspace about half of the time, which results in an increased probability of the ancillary qubit to be in the excited state after the $ZZ$ parity check and, thus, an increase of associated relaxation errors. A possible cause for the additional decrease in measured fidelity from cycle to cycle compared to the simulated one [Fig.~\ref{fig:pauli}(a)], is measurement-induced leakage of the ancillary qubit into its second excited state consistent with additional simulations we have performed.

In conclusion, we demonstrated the stabilization of a Bell state by repeated parity detection combined with conditional real-time feedback. More generally, our experiment demonstrates the use of projective stabilizer measurements to establish coherence between multiple qubits, without extracting information about the single-qubit states. Our comparison with Pauli frame updating provides evidence for potential advantages of real-time feedback control for quantum error correction by avoiding errors due to relaxation of the ancilla qubits, a topic to be further investigated. We find the measured steady state fidelity to be mainly limited by decoherence during the feedback delay time of $1\,\mu$s, which we expect to further decrease in the future by reducing the latency of feedback electronics~\cite{Salathe2018} and the readout duration~\cite{McClure2016, Bultink2016, Boutin2017, Walter2017}. Our results constitute an important step towards the real-time stabilization of entangled multi-qubit states beyond Bell states using higher-weight parity detection~\cite{Blumoff2016, Takita2016} as required, for example, in quantum error correction codes such as the Bacon-Shor code~\cite{Bacon2006} and the surface code~\cite{Raussendorf2007, Fowler2012}.

\section*{Acknowledgments}
The authors thank A.~Blais, B.~Royer, J.~M.~Renes  and J.~Home for valuable feedback on the manuscript and J.~Butscher, F.~Bruckmaier and M.~Bild for contributions to the experimental setup and the control software.

The authors acknowledge financial support by the Office of the Director of National Intelligence (ODNI), Intelligence Advanced Research Projects Activity (IARPA), via the U.S. Army Research Office grant W911NF-16-1-0071, by the National Centre of Competence in Research Quantum Science and Technology (NCCR QSIT), a research instrument of the Swiss National Science Foundation (SNSF), the EU Flagship on Quantum Technology  H2020-FETFLAG-2018-03 project 820363 OpenSuperQ and by ETH Zurich. The views and conclusions contained herein are those of the authors and should not be interpreted as necessarily representing the official policies or endorsements, either expressed or implied, of the ODNI, IARPA, or the U.S. Government.

\appendix

\section{Sample design, fabrication and characterization}
\label{app:sample}

\begin{figure}[tb]
\includegraphics[width=\linewidth]{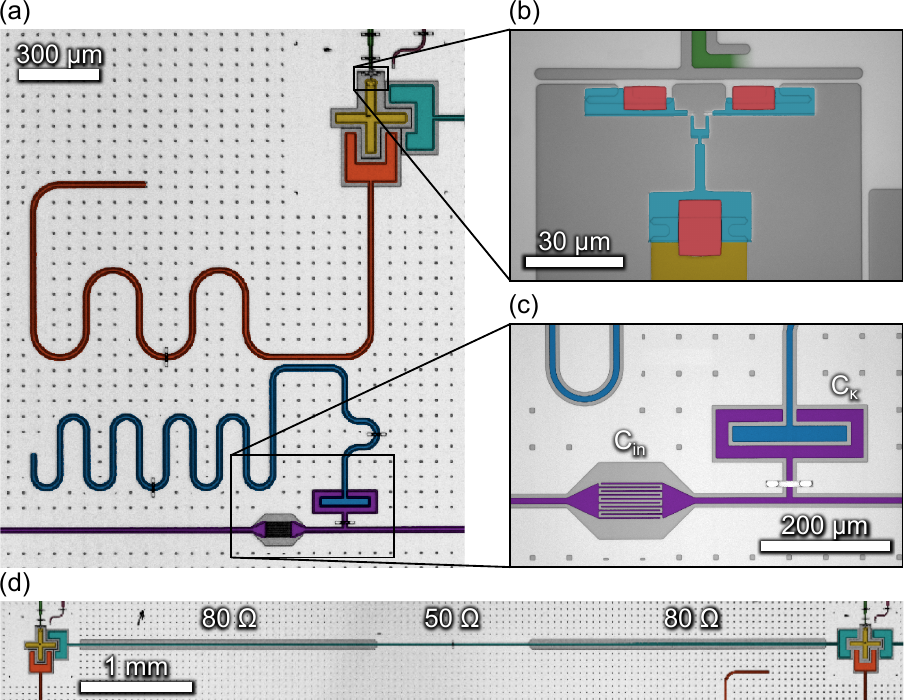}
\caption{False color optical micrographs of characteristic elements of the device.
(a) Data qubit $D1$ (yellow) coupled to its readout resonator (red). The readout resonator is coupled to the feedline (purple) through a Purcell filter (blue).
(b) Close-up of the SQUID (blue) contacted to the ground plane (light gray) and the qubit capacitor (yellow) using aluminum bandages (red).
(c) Input capacitor $C_{\rm in}$ of the feedline and coupling capacitor $C_\kappa$ to one of the Purcell filters.
(d) Coupling resonator (cyan) consisting of three segments with indicated impedances.}
\label{fig:special}
\end{figure}

The device in Fig.~\ref{fig:2} of the main text consists of four qubits coupled to each other in a linear chain.
All resonators, coupling capacitors, control lines and qubit islands are fabricated from a 150$\,$nm thin niobium film sputtered onto a high-resistivity intrinsic silicon substrate, which is patterned using photolithography and reactive ion etching. We further add airbridges to the device to establish a well-connected ground plane across the chip. The Josephson junctions of the qubits are fabricated using electron beam lithography and shadow evaporation of aluminum. As shown in Fig.~\ref{fig:special}(b), the junctions are arranged in a SQUID loop (blue) and contacted to the niobium film (gray and yellow) using an additional bandage of aluminum (red)~\cite{Dunsworth2017, Nersisyan2019}.

The qubits are connected via a readout resonator and a Purcell filter to a common feedline used for frequency multiplexed readout, see Fig.~\ref{fig:special}(a) and Ref.~\cite{Heinsoo2018} for details of this readout architecture. The capacitive coupling elements between the Purcell filters and the feedline, see Fig.~\ref{fig:special}(c), are designed to have a larger minimal feature size of $6\,\mu$m compared to the $3\,\mu$m of our standard interdigitated capacitors making their capacitance less sensitive to slight variations in the photo-lithography process. The qubits are designed to have a charging energy of $270$~MHz set by the total capacitance of the qubit island. An asymmetric SQUID reduces their sensitivity to flux noise~\cite{Hutchings2017}. From a fit to the measured qubit frequencies as a function of magnetic flux bias, we extract a SQUID asymmetry of approximately 1:8 for all three qubits. To achieve a large mutual inductance of $M\approx 1.6$~pH, we place the SQUID close to the shorted end of the flux line.

Qubit-qubit interactions are mediated by transmission line resonators, see Fig.~\ref{fig:special}(c). Since the coupling strength $J$ is proportional to the impedance of the coupling resonator~\cite{Majer2007, Koch2007}, we increase the characteristic impedance for segments of the transmission line resonator to $Z_0 \approx 80\,\Omega$ compared to the standard value of $50\,\Omega$. We achieve this increase in impedance by increasing the separation between center conductor and ground plane, which results in a smaller capacitance per unit length. The resulting qubit-qubit exchange rate is measured to be $J_{10\leftrightarrow 01}/2\pi \approx 3.8\,(3.4)~\text{MHz}$ at the interaction point of the two-qubit gates between qubits $D1$ ($D2$) and $A$. Additionally, we characterized the qubit parameters using standard spectroscopy and time domain measurements, see Table~\ref{tab:qb_params}.

\begin{table}[t]
\centering
\begin{tabular}{lccc}
\hline
\noalign{\vskip 1mm}
 & D1 & A & D2 \\
 \hline
 \hline
Qubit frequency, $\omega_q/2\pi$ (GHz) & 5.721 & 5.210 & 4.880 \\
Lifetime, $T_1$ ($\mu$s) & 19.7 & 13.7 & 23.4 \\
Ramsey decay time, $T_2^*$ ($\mu$s) & 12.5 & 11.7 & 11.2 \\
Echo decay time, $T_2^E$ ($\mu$s) & 22.4 & 14.5 & 15.0 \\
Readout frequency, $\omega_r/2\pi$ (GHz) & 6.892 & 7.087 & 6.687 \\
Readout linewidth, $\kappa_{eff}/2\pi$ (MHz) & 3.0 & 2.1 & 1.7 \\
Purcell filter linewidth, $\kappa_{P}/2\pi$ (MHz) & 27.2 & 34.7 & 10.7 \\
Purcell-readout coupling, $J_{PR}/2\pi$ (MHz) & 10.9 & 8.2 & 9.5 \\
Purcell-readout detuning, $\Delta_{PR}/2\pi$ (MHz) & 29.5 & 27.5 & 19.4 \\
Dispersive shift, $\chi/2\pi$ (MHz) & -3.9 & -1.6 & -1.8 \\
Thermal population, $P_{th}$ ($\%$) & 0.9 & 1.4 & 1.4 \\
Individual readout assingment prob. (\%) & 99.2 & 98.7 & 99.1 \\
Multiplexed readout assignment prob. (\%) & 98.7 & 98.9 & 99.1 \\
\hline
\end{tabular}
\caption{Measured sample parameters for the three qubits $D1$, $A$, and $D2$ used in the experiment.}
\label{tab:qb_params}
\end{table}

\section{Experimental Setup and Timing Diagram}
\label{app:setup}

We mount our device, shown in Fig~\ref{fig:2}, at the base temperature plate of a dilution refrigerator, where the device is protected from ambient magnetic fields by $\mu$-metal and aluminum shields.

\begin{figure*}[bt]
\includegraphics[width=\linewidth]{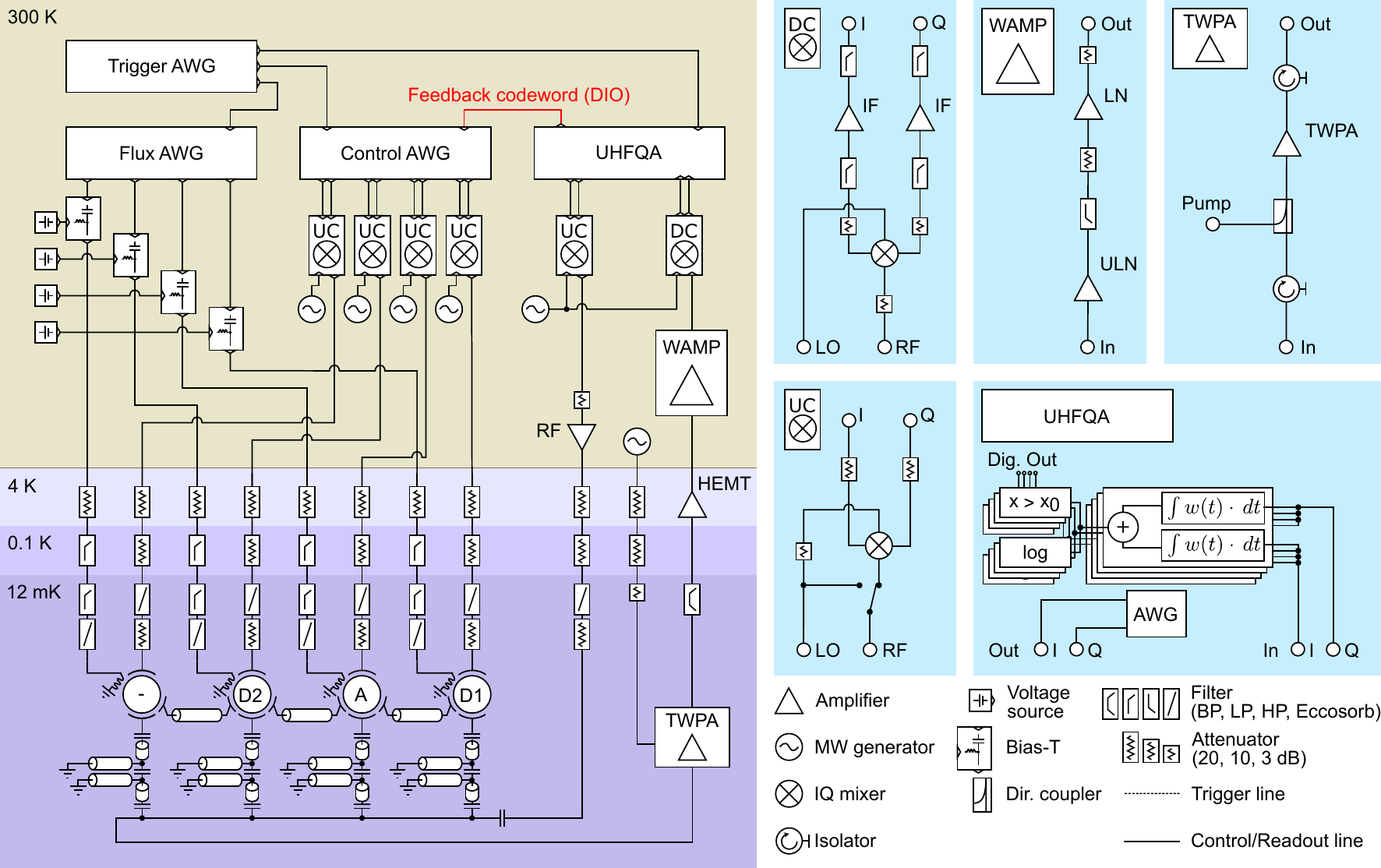}
\caption{Schematics of the experimental setup and the filtering of the input and output signals including low-pass (LP), band-pass (BP) and infrared-blocking Eccosorb filters and attenuators at the indicated temperature stages. Extended schematics of the components of the readout chain are shown in the panels. The readout signal is amplified using a travelling-wave parametric amplifier (TWPA), a high-electron-mobility transistor (HEMT) amplifier, and the warm amplifier board (WAMP) consisting of an ultra-low-noise (ULN) amplifier, a low-noise (LN) amplifier and a high-pass (HP) filter. The UHFQA supports weighted integration, data logging (log) and thresholding ($x>x_0$) of the readout signal. Further details are given in the text.} \label{fig:detailedsetup}
\end{figure*}

\begin{figure}[t]
\includegraphics[width=\linewidth]{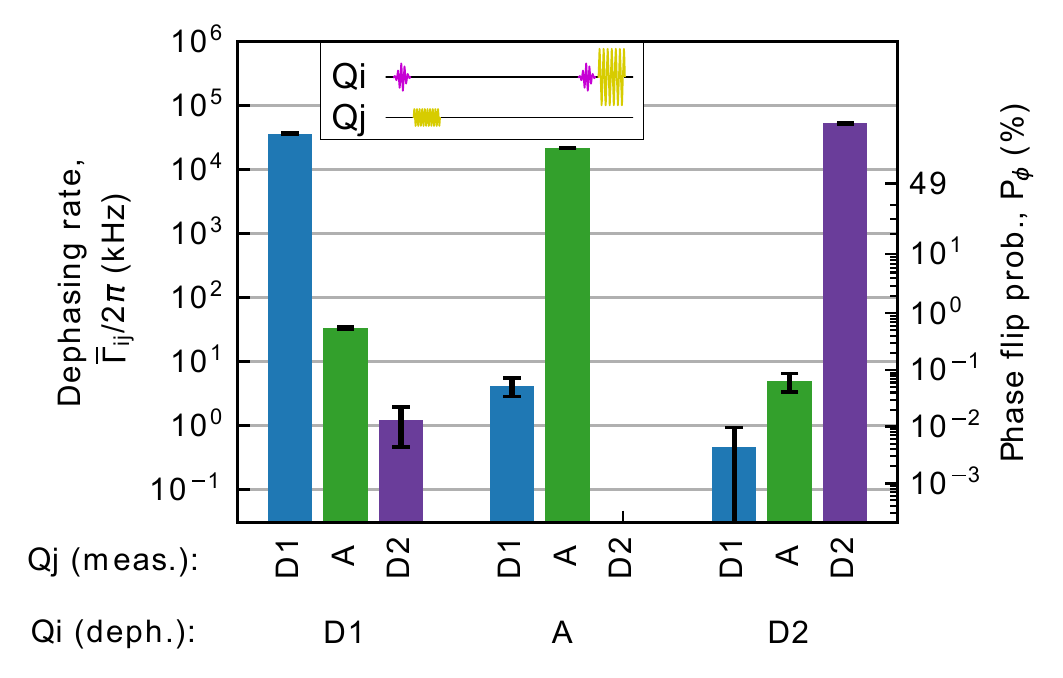}
\caption{Average dephasing rate (left axis) of qubit $Qi$ during a readout pulse on qubit $Qj$ and the corresponding probability of a phase error (right axis) on qubit $Qi$. The inset shows the pulse sequence with $\pi/2$-rotations in purple and readout pulses in yellow.} \label{fig:measdephase}
\end{figure}

Each qubit is coupled to a charge control line used for applying microwave pulses for single qubit rotations and a flux line used for tuning the qubit frequencies in-situ to realize two-qubit gates, see Fig.~\ref{fig:detailedsetup}. The flux pulses are generated using an arbitrary waveform generator (AWG, \emph{Tektronix 5014c}) with a sampling rate of 1.2~GSa/s and combined with a constant DC current using a bias-tee, before routing the signal through a chain of attenuators and low-pass filters to the sample. We use eccosorb filters to attenuate infrared radiation.
The  baseband microwave control pulses are generated by the control AWG (\emph{Zurich Instruments HDAWG}) with 8 channels and a sampling rate of 2.4~GSa/s at an intermediate frequency (IF) of $100$~MHz.  The baseband pulses are upconverted to microwave frequencies using IQ mixers installed on upconversion boards (UC). The upconverted microwave pulses are routed from room temperature to the quantum device through a chain of 20~dB attenuators at the 4~K, 100~mK and 12~mK stages.
We perform multiplexed readout by probing the feedline of the device with a readout pulse which has frequency components at each readout resonance frequency~\cite{Heinsoo2018}. The readout pulses are generated and detected using  an FPGA based control system (\emph{Zurich Instruments UHFQA}) with a sampling rate of 1.8~GSa/s. The UHFQA outputs a probe-pulse, which is upconverted and transmitted to the sample through lines similar to the ones used for the drive pulses. The measurement signal picked up at the output of the sample is amplified using a wide bandwidth near-quantum-limited traveling wave parametric amplifier (TWPA)~\cite{Macklin2015} connected to wideband 3-12 GHz isolators at its input and output. Moreover, we installed a bandpass filter in the output line to suppress amplifier noise outside the bandwidth of the isolators of our detection chain. The signal is further amplified by a high-electron-mobility transistor (HEMT) amplifier at the 4~K stage and amplifiers at room temperature (WAMP). Finally, the signal is downconverted (DC) and then processed using the weighted integration units of the UHFQA.
By comparing the single shot readout SNR with the measurement induced dephasing rate, see Fig.~\ref{fig:measdephase}, we extract an overall quantum efficiency of the detection chain of $\eta=24\%$~\cite{Bultink2018, Heinsoo2018}.

We use a dedicated trigger AWG (\emph{Tektronix 5014c}) to synchronize our instruments. We program the trigger AWG marker channels to trigger each readout pulse.
The control AWG is triggered at the beginning of each parity stabilization round and a separate trigger is used to initiate the feedback pulses. At the arrival of the feedback trigger, the control AWG generates a pulse conditioned on the latest readout result, which is communicated by the UHFQA to the control AWG via a digital input-output (DIO) line.
The feedback delay of the experiment is determined by the readout integration time of 400~ns and an electronic delay of 600~ns, which is dominated by internal delays of the UHFQA and the control AWG.
The pulses on the flux AWG are precompiled into a single waveform such that we can apply an infinite impulse response (IIR) filter to compensate for the frequency dependent response of the flux line.

\begin{figure}[tb]
\includegraphics[width=\linewidth]{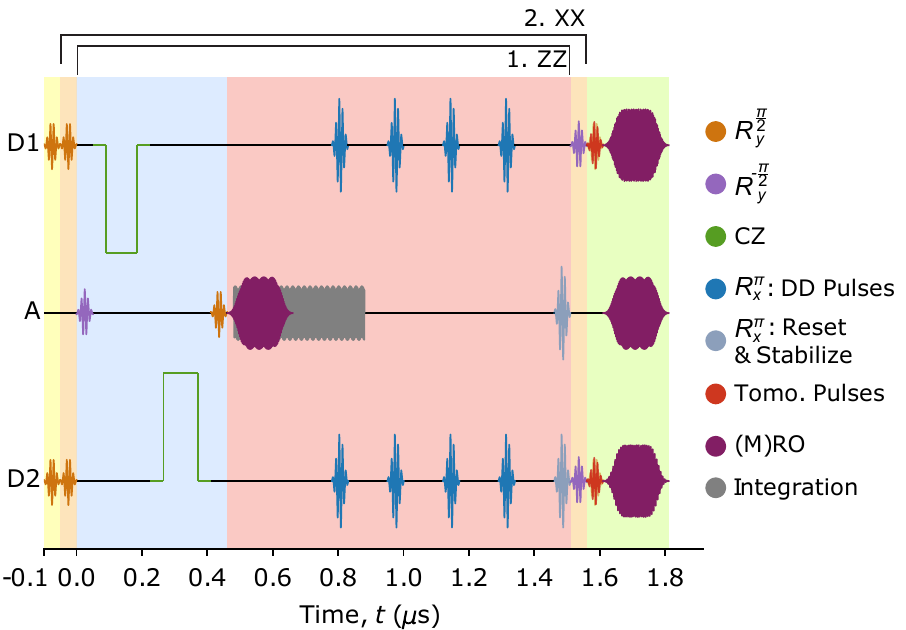}
\caption{Pulse sequence for the repeated parity stabilization experiment with $\pi/2$-rotations around the $Y$-axis in brown, $-\pi/2$-rotations around the $Y$-axis in light purple, two-qubit gates in green, $\pi$-rotations around the $X$-axis for dynamical decoupling in blue, conditional feedback $\pi$-rotations about the $X$-axis in light blue, tomography pulses in red and readout pulses in purple. The integration window for the ancilla readout is shown in gray.} \label{fig:pulse}
\end{figure}

The waveforms generated by the classical control hardware are shown in Fig.~\ref{fig:pulse} and implement each stabilization cycle in a total time of $1.51\,\mu$s. Prior to these waveforms is an additional multiplexed readout pulse (not shown) used for heralding the qubit initial state to be ground state of all qubits~\cite{Heinsoo2018}. All single qubit gates are realized as DRAG pulses~\cite{Motzoi2009}, to avoid phase errors and leakage due to the presence of the second excited transmon state. The pulses are implemented with a Gaussian envelope truncated to $5\sigma$ with $\sigma=10$~ns such that the total pulse duration is 50 ns. The flux pulses have lengths of 96 ns and 105 ns for $D1$ and $D2$ respectively.
Before and after each flux pulse, we insert a buffer of 40 ns in order to avoid overlap of single qubit pulses with the rising and falling edges of the flux pulses.
We compensate for distortions of the flux pulses due to the bias-tee and the frequency-dependent response of the flux line with IIR and FIR (finite impulse response) filters to achieve the desired pulse shape at the quantum device.
The readout pulses are 200 ns long square pulses convolved with a Gaussian kernel with a standard deviation of $\sigma = 20$~ns for reduced cross-dephasing, see Appendix~\ref{app:readout}. During the feedback delay in the experiment, we apply 4 Carr-Purcell-Meiboom-Gill (CPMG) dynamical decoupling pulses, which (ideally) cancel the effect of residual $ZZ$-coupling and increase the dephasing time of the qubits.

\begin{figure}[t]
\includegraphics[width=\linewidth]{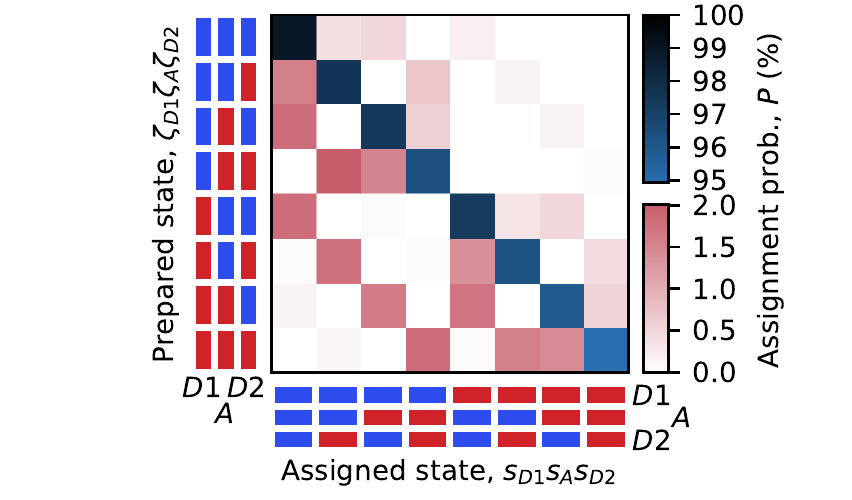}
\caption{
Probability of assigning the 3-qubit state $s_{D1}s_{A}s_{D2}$ when preparing state $\zeta_{D1}\zeta_{A}\zeta_{D2}$. Ground and excited states are labeled blue and red, respectively.} \label{fig:probmat}
\end{figure}

\section{Readout Characterization}
\label{app:readout}
With the dispersive shifts $\chi$ and the resonator linewidths $\kappa_{\rm eff}$ listed in Table~\ref{tab:qb_params} of Appendix~\ref{app:sample}, we measure correct readout assignment probabilities of approximately 99\%, see Table~\ref{tab:qb_params}, for both individual and multiplexed readout, the agreement of which indicates low readout cross-talk~\cite{Heinsoo2018}. More specifically, we find probability to correctly assign the state of any initial multi-qubit state of more than $95\%$, see Fig.~\ref{fig:probmat}. We also extract the average cross-measurement induced dephasing rate, $\bar{\Gamma}_{ij}$ from the loss of contrast in the Ramsey signal of qubit $Qi$ when interleaving a readout pulse on qubit $Qj$ between two Ramsey pulses~\cite{Heinsoo2018, Bultink2018}. We find that the probability of inducing a phase error on the data qubits due to the ancilla readout pulse is below $0.3\%$, see Fig.~\ref{fig:measdephase}. We observe, however, that the readout of qubit $A$ induces a deterministic phase shift on qubits $D1$ and $D2$ of $33.4^{\circ}$ and $33.2^{\circ}$ respectively, which we attribute to measurement induced Stark shifts from off-resonantly driving the coupling resonators. In the parity stabilization protocol, we compensate for these phase shifts using virtual $Z$ gates~\cite{McKay2017}.

\section{Master equation simulations}
\label{app:simulations}

To understand the physical origin of the  reduced fidelities observed in our experiments, we perform numerical simulations of the experimental protocol including a set of error sources, which we were able to identify. We simulate the time evolution of the system Hamiltonian by solving a master equation, while the ancilla measurements are modelled using the positive-operator valued measure (POVM) formalism~\cite{Wiseman2010}.

The master equation modeling the time-evolution is given by
\begin{align}
\dot{\rho} = -\frac{i}{\hbar} [ H(t), \rho ]  + \sum_i \Big[ \hat{c}_i \rho \hat{c}_i\dag - \frac{1}{2} \Big( \hat{c}_i\dag\hat{c}_i \rho + \rho \hat{c}_i\dag \hat{c}_i \Big) \Big], \label{eq:mastereq}
\end{align}
where $\rho$ is the density matrix describing the system at time $t$ and $H(t)$ is the Hamiltonian, the time-dependence of which models the applied gate sequence. The collapse operators $\hat{c}_i$ model incoherent processes. We solve the master equation numerically using the software package QuTIP version 4.2~\cite{Johansson2013a}.

To simplify the description of the system's time evolution, we consider the Hamiltonian to be piece-wise constant. For example, we simulate the preparation pulses, i.e., two $\pi/2$ pulses on D1 and D2, using the Hamiltonian
\begin{align*}
H/\hbar = \alpha_{\pi/2} \frac{\sigma_y}{2} \otimes \mathbb{I} \otimes \mathbb{I} + \alpha_{\pi/2} \mathbb{I} \otimes \mathbb{I} \otimes \frac{\sigma_y}{2}
\end{align*}
for a duration $t_g = 50$~ns with the amplitude $\alpha_\theta = \theta/t_g$. Here, the Hamiltonian is expressed in the basis $D1\otimes A \otimes D2$. Similarly, we simulate the controlled phase gate, e.g., between $D1$ and $A$, by evolving $\rho$ according to the Hamiltonian $H/\hbar = (\pi/t_{fp}) \ket{11}\bra{11} \otimes \mathbb{I}$, where $t_{fp}$ is the length of the flux pulse. While this method of simulating the controlled phase gate generates the ideal coherent evolution, it does not include leakage into the $\ket{02}$-state.
Moreover, we include a buffer time of 40~ns before and after the flux pulse with $H=0$. In addition, we include the constant Hamiltonian
\begin{align*}
H_{ZZ}/\hbar &= j_{D1A} \ket{11}\bra{11} \otimes \mathbb{I} + j_{D2A} \mathbb{I} \otimes \ket{11}\bra{11}
\end{align*}
to account for the residual $ZZ$-coupling between the qubits. Here, we use $j_{D1A}/2\pi = 110$~kHz and $j_{D2A}/2\pi = 370$~kHz, which we have measured independently in a Ramsey experiment.
The incoherent errors are described by the Lindblad terms in Eq.~\eqref{eq:mastereq}. In particular, we use the collapse operators
\begin{align*}
\hat{c}_{T_1,i} &= \sqrt{\frac{1}{T_{1,i}}} \sigma_{-,i}, \\
\hat{c}_{T_{\phi,i}} &=  \sqrt{ \frac{1}{2} \Big( \frac{1}{T_{2,i}} - \frac{1}{2T_{1,i}} \Big) } \sigma_{z,i},
\end{align*}
where $T_{1,i}$ and $T_{2,i}$ are the lifetime and decoherence time of qubit $i$, respectively.

To simulate the ancilla measurement, we consider the POVM operators:
\begin{align}
M_1 &= \sqrt{P(0|0)} \ket{0}\bra{0}_A + \sqrt{P(0|1)} \ket{1}\bra{1}_A, \\
M_{-1} &= \sqrt{P(1|0)} \ket{0}\bra{0}_A + \sqrt{P(1|1)} \ket{1}\bra{1}_A,
\end{align}
for the outcomes 1 and $-1$ respectively, where $P(i|j)$ are the probabilities for measuring the state $i$ when preparing the state $j$. Here, for simplicity, we choose POVM operators corresponding to a minimal disturbance measurement~\cite{Wiseman2010}.  We extract the $P(i|j)$ from the measured single-shot readout histograms and find them to be $P(0|0) = 99.4\%$, $P(1|0) = 0.54\%$, $P(0|1) = 2.1\%$ and $P(1|1) = 97.95\%$.

\begin{figure}[b]
\includegraphics[width=\linewidth]{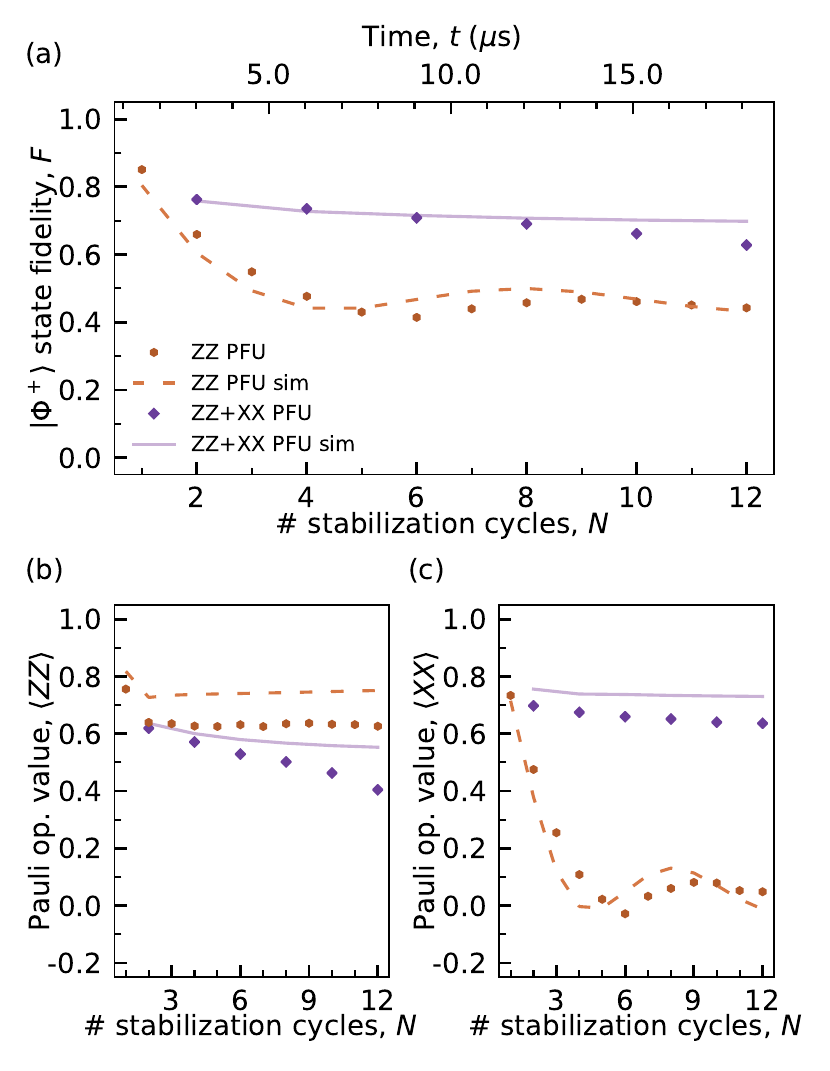}
\caption{
(a) Measured (dots and squares) and simulated (solid and dashed lines) Bell state fidelity after $N$ rounds of $ZZ$ parity stabilization (brown), and when iterating between $ZZ$ and $XX$ (purple) using Pauli frame updates (PFU) of the data qubits rather than real-time feedback. (b) - (c) Measured and simulated expectation values $\langle ZZ \rangle$ and $\langle XX\rangle$ for the same experiments.
}
\label{fig:pauli}
\end{figure}

We evaluate the probability for each ancilla measurement outcome as $p_i = \text{Tr}(M_i \rho(t_m) M_i\dag)$ for $i=\pm 1$, where $\rho(t_m)$ is the density matrix at the time of measurement $t_m$. We describe the density matrix conditioned on the measurement outcome by the density matrix $\rho_i(t_m) = M_i \rho(t_m) M_i\dag / p_i$~\cite{Wiseman2010}. We keep track of the time-evolution for
both possible states $\rho_{\pm 1 }$ and simulate their respective time-evolution during the feedback delay time of $t_d = 1.0\,\mu$s, during which we apply the 4 CMPG dynamical decoupling pulses explicitly in the simulations.
After the time $t_d$, the feedback pulse is applied to the state $\rho_{-1}$. As for all the single qubit gates, the feedback pulse has a duration $t_g = 50$~ns. On the other hand, $\rho_1$ does not receive any feedback and is evolved for a time $t_g$ with no control Hamiltonian applied. We combine the two density matrices at the time $t_f = t_m + t_d + t_g$ to obtain the unconditional density matrix
\begin{align}
\rho(t_f) =&\; p_1 \rho_1(t_f) + p_{-1} \rho_{-1}(t_f),
\end{align}
at the end of the parity stabilization cycle.

\section{Pauli Frame Updating}
\label{app:pauli}

As an alternative to the active stabilization protocol presented in the main text, we may choose to keep track of the parity measurements in software and apply Pauli frame updating to stabilize the target subspaces. The gate sequence is then equivalent to Fig.~\ref{fig:1}, but feedback is only used for resetting the ancilla qubit after each parity measurement. As the ancilla is reset in every stabilization round, the Pauli frame update is only conditioned on the last two (one) parity measurements when stabilizing both $ZZ$ and $XX$ (only $ZZ$). Thus, if the last $ZZ$ ($XX$) outcome is $-1$, we apply a $R_x^\pi$ ($R_z^\pi$) rotation to the Pauli operators of the state tomography of $D2$ before averaging the data and reconstructing the most likely density matrix.

In Fig.~\ref{fig:pauli}, we observe that the initial state fidelity is slightly above the one of the active stabilization protocol in Fig.~\ref{fig:4}, however, the fidelity now decreases when repeating the protocol. This decreasing fidelity is partly expected from simulations due to the data qubits not being actively stabilized in the even subspaces of the parity operators which leads to the ancilla qubit being in the $\ket{1}$-state more often. Thus, during the feedback delay, there is a higher chance of $T_1$-errors on the ancilla qubit, which will propagate into the next parity measurement cycle. We also expect that the residual $ZZ$ errors, when the ancilla is in the excited state, leads to accumulated phase errors in $\exv{XX}$ when only stabilizing $ZZ$, see Fig.~\ref{fig:pauli}(c). These accumulated errors are observed in the data and are reproduced in the simulations. Beyond the errors predicted by the simulations, we observe an additional decay of the $\langle ZZ \rangle$ correlations. While these errors are significantly larger than expected from simulations, they could be explained by measurement-induced transitions from the $\ket{1}$-state to the second-excited state of the ancilla qubit during readout.

\bibliography{QudevRefDB}
\end{document}